\documentclass[sigconf]{acmart}

\usepackage{balance}
\usepackage{graphicx}
\usepackage{epstopdf}
\usepackage{booktabs} 
 \usepackage{amsmath}

\usepackage{amsmath}
\usepackage{bm}
\usepackage{algorithm}
\usepackage{algorithmic}
\usepackage{amsmath}
\usepackage{multirow}
\usepackage{indentfirst}
\usepackage{subfigure}
\usepackage{epsfig}
\usepackage{epstopdf}
\usepackage{xspace}

\usepackage{subeqnarray}

\newcommand{\paratitle}[1]{\vspace{1.5ex}\noindent\textbf{#1}}
\newcommand{\ie}{\emph{i.e.,}\xspace}

\newcommand{\eg}{\emph{e.g.,}\xspace}
\newcommand{\etal}{\emph{et al.}\xspace}

\newcommand{\name}{AlphaStock\xspace}
\newcommand{\ignore}[1]{}

\newtheorem{Def}{Definition}
\def\BibTeX{{\rm B\kern-.05em{\sc i\kern-.025em b}\kern-.08emT\kern-.1667em\lower.7ex\hbox{E}\kern-.125emX}}

\begin{document}

\copyrightyear{2019}
\acmYear{2019}
\setcopyright{acmcopyright}
\acmConference[KDD '19]{The 25th ACM SIGKDD Conference on Knowledge Discovery and Data Mining}{August 4--8, 2019}{Anchorage, AK, USA}
\acmBooktitle{The 25th ACM SIGKDD Conference on Knowledge Discovery and Data Mining (KDD '19), August 4--8, 2019, Anchorage, AK, USA}
\acmPrice{15.00}
\acmDOI{10.1145/3292500.3330647}
\acmISBN{978-1-4503-6201-6/19/08}

\title{AlphaStock: A Buying-Winners-and-Selling-Losers Investment Strategy using Interpretable Deep Reinforcement Attention Networks}

\author{Jingyuan Wang$^{1,4}$, Yang Zhang$^1$, Ke Tang$^2$, Junjie Wu$^{3,4,\ast}$, Zhang Xiong$^1$}
\affiliation{%
  \institution{$1.$MOE Engineering Research Center of Advanced Computer Application Technology, \\ School of Computer Science Engineering, Beihang University, Beijing, China \\
  $2.$Institute of Economics, School of Social Sciences, Tsinghua University, Beijing China \\
  $3.$Beijing Key Laboratory of Emergency Support Simulation Technologies for City Operations, \\ School of Economics and Management, Beihang University, Beijing, China \\
  $4.$Beijing Advanced Innovation Center for BDBC, Beihang University, Beijing, China. $\ast$ Corresponding author.}}

%
\renewcommand{\shortauthors}{J. Wang, et al.}

%
\begin{abstract}
Recent years have witnessed the successful marriage of finance innovations and AI techniques in various finance applications including quantitative trading (QT). Despite great research efforts devoted to leveraging deep learning (DL) methods for building better QT strategies, existing studies still face serious challenges especially from the side of finance, such as the balance of risk and return, the resistance to extreme loss, and the interpretability of strategies, which limit the application of DL-based strategies in real-life financial markets. In this work, we propose \emph{\name}, a novel reinforcement learning (RL) based investment strategy enhanced by interpretable deep attention networks, to address the above challenges. Our main contributions are summarized as follows: $i$) We integrate deep attention networks with a Sharpe ratio-oriented reinforcement learning framework to achieve a risk-return balanced investment strategy; $ii$) We suggest modeling interrelationships among assets to avoid selection bias and develop a cross-asset attention mechanism; $iii$) To our best knowledge, this work is among the first to offer an interpretable investment strategy using deep reinforcement learning models. The experiments on long-periodic U.S. and Chinese markets demonstrate the effectiveness and robustness of \name over diverse market states. It turns out that \name tends to select the stocks as winners with high long-term growth, low volatility, high intrinsic value, and being undervalued recently. 
\end{abstract}

%
%


%

\maketitle

{\fontsize{8pt}{8pt} \selectfont
\noindent \textbf{ACM Reference Format:}\\
Jingyuan Wang, Yang Zhang, Ke Tang, Junjie Wu, Zhang Xiong. 2019. AlphaStock: A Buying-Winners-and-Selling-Losers Investment Strategy using Interpretable Deep Reinforcement Attention Networks In {\it The 25th ACM SIGKDD Conference on Knowledge Discovery Data Mining (KDD '19), August 4--8, 2019, Anchorage, AK, USA.} ACM, NY, NY, USA, 9 pages.\\ https://doi.org/10.1145/3292500.3330647 }

\section{Introduction}
\label{sec:intro}

Given the ability in handling large scales of transactions and offering rational decision-makings, quantitative trading (QT) strategies have long been adopted in financial institutions and hedge funds and have achieved spectacular successes.
Traditional QT strategies are usually based on specific financial logics. For instance, the {\em momentum} phenomenon found by Jegadeesh and Titman in the stock market~\cite{bwsl} was used to build momentum strategies. The {\em mean reversion}~\cite{poterba1988mean} proposed by Poterba and Summers believes that asset price tends to move to the average over time, so the bias of asset prices to their means could be used to select investment targets. The {\em multi-factor} strategy~\cite{fama1996multifactor} uses factor-based asset valuations to select assets. Most of these traditional QT strategies, though equipped with solid financial theories, can only leverage some specific characteristic of financial markets, and therefore might be vulnerable to complex markets with diverse states.

In recent years, deep learning (DL) emerges as an effective way to extract multi-aspect characteristics from complex financial signals. Many supervised deep neural networks are proposed in the literature to predict asset prices using various factors, such as frequency of prices~\cite{hu2017state}, economic news~\cite{news1}, social media~\cite{tweets}, and financial events~\cite{event1,event2}. Deep neural networks are also adopted in reinforcement learning (RL) frameworks to enhance traditional shallow investment strategies~\cite{dql_stanford,deng2017deep,ding2018investor}. Despite the rich studies above, applying DL to real-life financial markets still faces several challenges:


{\it Challenge 1: Balancing return and risk.} Most existing supervised deep learning models in finance focus on price prediction without risk awareness, which is not in line with fundamental investment principles and may lead to suboptimal performance~\cite{fischer2018reinforcement}. While some RL-based strategies~\cite{fischer2018reinforcement,moody1998performance} have considered this problem, how to adopt state-of-the-art DL approaches into risk-return-balanced RL frameworks, is yet not well studied.

{\it Challenge 2: Modeling interrelationships among assets.} Many financial tools in the market can be used to derive risk-aware profits from the interrelationship among assets, such as hedging, arbitrage, and the BWSL strategy used in this work. However, existing DL/RL-based investment strategies paid little attention to this important information.


{\it Challenge 3: Interpreting investment strategies.} There is a longstanding voice arguing that DL-based systems are ``unexplainable black boxes'' and therefore cannot be used in crucial applications like medicine, investment and military~\cite{guidotti2018survey}. RL-based strategies with deep structures make it even worse. How to extract interpretable rules from DL-enabled strategies remains an open problem.


In this paper, we propose \emph{\name}, a novel reinforcement learning based strategy using deep attention networks, to overcome the above challenges. \name is essentially a {\em buying winners and selling losers} (BWSL) strategy for stock assets. It consists of three components. The first is a {\em Long Short-Term Memory with History state Attention} (LSTM-HA) network, which is used to extract asset representations from multiple time series. The second component is a {\em Cross-Asset Attention Network} (CAAN), which can fully model the interrelationships among assets as well as the asset price rising prior. The third is a portfolio generator, which gives the investment proportion of each asset according to the output winner scores of the attention networks. We use a RL framework to optimize our model towards a return-risk-balanced objective, \ie maximizing the Sharpe Ratio. In this way, the merit of representation learning via deep attention models and the merit of risk-return balance via Sharpe ratio targeted reinforcement learning are integrated naturally. Moreover, to gain interpretability for \name, we propose a sensitivity analysis method to unveil how our model selects an asset to invest according to its multi-aspect features.

Extensive experiments on long-periodic U.S. stock markets demonstrate that our \name strategy outperforms some state-of-the-art competitors in terms of a variety of evaluation measures. In particular, \name shows excellent adaptability to diverse market states (enabled by RL and Sharpe ratio) and exceptional ability for extreme loss control (enabled by CAAN). Extended experiments on Chinese stock markets further confirm the superiority of \name and its robustness. Interestingly, the interpretation analysis results reveal that \name selects assets by following a principle as ``selecting the stocks as winners with high long-term growth, low volatility, high intrinsic value, and being undervalued recently''. 

\section{Preliminaries}\label{sec:preliminaries}

In this section, we first introduce the financial concepts used throughout this paper, and then formally define our problem.

\subsection{Basic Financial Concepts}

\begin{Def}[Holding Period]
A holding period is a minimum time unit to invest an asset. We divide the time axis as sequential holding periods with fixed length, such as one day or one month. We call the starting time of the $t$-th holding period as the time $t$.
\end{Def}

\begin{Def}[Sequential Investment]
A sequential investment is a sequence of holding periods. For the $t$-th holding period, a strategy uses original capital to invest in assets at time $t$, and gets profits (could be negative) at time $t+1$. The capitals plus profits of the $t$-th holding period are used as the original capitals of the $(t+1)$-th holding period.
\end{Def}


\begin{Def}[Asset Price] The price of an asset is defined as a time series $\bm{p}^{(i)} = \{p_1^{(i)}, p_2^{(i)}, \ldots, p_t^{(i)}, \ldots\}$, where $p_t^{(i)}$ denotes the price of asset $i$ at time $t$.
\end{Def}

In this work, we use a stock as an asset to describe our model, which could be extended to other types of assets by taking asset specificities and transaction rules into consideration.

\begin{Def}[Long Position]
The long position is the trading operation that buys an asset at time $t_1$ first and then sells it at $t_2$. The profit of a long position during the period from $t_1$ to $t_2$ for asset $i$ is $u_i(p_{t_2}^{(i)} - p_{t_1}^{(i)})$, where $u_i$ is the buying volume of asset $i$.
\end{Def}

In the long position, traders expect an asset will rise in price, so they buy the asset first and wait for the price rise to earn profits.

\begin{Def}[Short Position]
A short position is the trading operation that sells an asset at $t_1$ first and then buys it back at $t_2$. The profit of a short position during the period from $t_1$ to $t_2$ for asset $i$ is $u_i(p_{t_1}^{(i)} - p_{t_2}^{(i)})$, where $u_i$ is the selling volume of asset $i$.
\end{Def}

Short position is a reverse operation of the long position. Traders' expectation in short position is that the price will drop, so they sell at a price higher than the price at which they buy it back later. In the stock market, a short position trader borrows stocks from a broker and sells them at $t_1$. At $t_2$, the trader buys the sold stocks back and returns them to the broker.

\begin{Def}[Portfolio]
Given an asset pool with $I$ assets, a portfolio is defined as a vector $\bm{b} = (b^{(1)},$ $\ldots,$ $b^{(i)},$ $\ldots,$ $b^{(I)})^{\top}$, where $b^{(i)}$ is the proportion of the investment on asset $i$, with $\sum_{i=1}^I b^{(i)} = 1$. 
\end{Def}


Assume we have a collection of portfolios $\{\bm{b}^{(1)},\ldots,\bm{b}^{(j)},\ldots,\bm{b}^{(J)}\}$. The investment on portfolio $\bm{b}^{(j)}$ is $M^{(j)}$, with $M^{(j)}\geq 0$ when taking a long position on $\bm{b}^{(j)}$, and $M^{(j)}\leq 0$ when taking a short position. We then have the following important definition.

\begin{Def}[Zero-investment Portfolio]
A zero-investment portfolio is a collection of portfolios that has a net total investment of zero when the portfolios are assembled. That is, for a zero-investment portfolio containing $J$ portfolios, the total investment $\sum_{j=1}^J M^{(j)}=0$. 
\end{Def}

For instance, an investor may borrow \$1,000 worth of stocks in one set of companies and sell them as a short position, and then use the proceeds of short selling to purchase \$1,000 stocks in another set of companies as a long position. The assemble of the long and short positions is a zero-investment portfolio. Note that while the name is ``zero-investment'', there still exists a budget constraint to limit the overall worth of stocks that can be borrowed from the broker. Also, we ignore real-world transaction costs for simplicity.

\subsection{The BWSL Strategy}\label{sec:define}


In this paper, we adopt the \emph{buy-winners-and-sell-losers} (BWSL) strategy for stock trading~\cite{bwsl}, the key of which is to buy the assets with high price rising rate (winners) and sell those with low price rising rate (losers). We execute the BWSL strategy as a zero-investment portfolio consisting of two portfolios: a long portfolio for buying winners and a short portfolio for selling losers. Given a sequential investment with $T$ periods, we denote the short portfolio for the $t$-th period as $\bm{b}^-_t$ and the long portfolio as $\bm{b}^+_t$, $t=1,\ldots,T$.

At time $t$, given a budget constraint $\tilde{M}$, we borrow the ``loser'' stocks from brokers according to the investment proportion in $\bm{b}^-_t$. The volume of stock $i$ that we can borrow is
\begin{equation}\label{}\small
  u_t^{-(i)} = {\tilde{M}\cdot b^{-(i)}_t}/{p_t^{(i)}},
\end{equation}
where $b^{-(i)}_t$ is the proportion of stock $i$ in $\bm{b}^-_t$. Next, we sell the ``loser'' stocks we borrowed and get the money $\tilde{M}$. After that, we use $\tilde{M}$ to buy the ``winner'' stocks according to the long portfolio $\bm{b}^+_t$. The volume of stock $i$ that we can buy at time $t$ is
\begin{equation}\label{}\small
  u_t^{+(i)} = {\tilde{M}\cdot b^{+(i)}_t}/{p_t^{(i)}}.
\end{equation}
The money $\tilde{M}$ we used to buy winner stocks is the proceeds of short selling, so the net investment on the portfolio $\{\bm{b}^+_t,\bm{b}^-_t\}$ is zero.

At the end of the $t$-th holding period, we sell stocks in the long portfolio. The money we can get is the proceeds of selling stocks using new prices at $t+1$ for all stocks, \ie
\begin{equation}\label{}\small
   M_t^{+} = \sum_{i=1}^I u_t^{+(i)} {p_{t+1}^{(i)}} = \sum_{i=1}^I \tilde{M}\cdot b^{+(i)}_t \frac{p_{t+1}^{(i)}}{p_t^{(i)}}.
\end{equation}
Next, we buy the stocks in the short portfolio back and return them to the broker. The money we spend on buying the short stocks is
\begin{equation}\label{}\small
  M_t^{-} = \sum_{i=1}^{I'} u_t^{-(i)} {p_{t+1}^{(i)}} = \sum_{i=1}^{I'} \tilde{M}\cdot b^{-(i)}_t \frac{p_{t+1}^{(i)}}{p_t^{(i)}}.
\end{equation}
The ensemble profit earned by the long and short portfolios is $M_t = M_t^{+} - M_t^{-}$. Let $z_{t}^{(i)}= {p_{t+1}^{(i)}}/{p_t^{(i)}}$ denote the {\em price rising rate} of stock $i$ in the $t$-th holding period. Then, the {\em rate of return} of the ensemble portfolio is calculated as
\begin{equation}\label{eq:return_rate}\small
  R_t = \frac{{M}_t}{\tilde{M}} = \sum_{i=1}^I b^{+(i)}_t z_{t}^{(i)} - \sum_{i=1}^{I'} b^{-(i)}_t z_{t}^{(i)}.
\end{equation}

\paratitle{Insight I.} As shown in Eq.~\eqref{eq:return_rate}, a positive profit, \ie $R_t > 0$, means the average price rising rate of stocks in the long portfolio is higher than that in the short portfolio, \ie
\begin{equation}\label{}\small
  \sum_{i=1}^I b^{+(i)}_t z_{t}^{(i)} > \sum_{i=1}^{I'} b^{-(i)}_t z_{t}^{(i)}.
\end{equation}
A profitable BWSL strategy must ensure the stocks in the portfolio $\bm{b}^{+}$ have a higher average price rising rate than the stocks in $\bm{b}^{-}$. That is to say, even the prices of all stocks in the market are falling, as long as we can ensure the price falling of stocks in $\bm{b}^{+}$ is slower than that in $\bm{b}^{-}$, we can still get profits. On the contrary, even the prices of all stocks are rising, if the rising of stocks in $\bm{b}^{-}$ is faster than that in $\bm{b}^{+}$, our strategy still lose money. This characteristic implies that the absolute price rising or falling of stocks is not the main concern of our strategy; rather, the relative price relations among stocks are much more important. As a consequence, we must design a mechanism to describe the interrelationships of stock prices in our model for the BWSL strategy.


\subsection{Optimization Objective}

In order to ensure that our strategy considers both return and risk of an investment, we adopt the {\em Sharpe ratio}, a risk-adjusted return developed by the Nobel laureate William F. Sharpe~\cite{sharpe} in 1994, to measure the performance of our strategy.
\begin{Def}[Sharpe Ratio]
The Sharpe ratio is the average return in excess of the risk-free return per unit of volatility. Given a sequential investment that contains $T$ holding periods, its Sharpe ratio is calculated as
\begin{equation}\label{eq:sharpe_ratio}\small
  H_T = \frac{A_T - \Theta}{V_T},
\end{equation}
where $A_T$ is the average rate of return per period for the investment, $V_T$ is the volatility
that is used to measure risk of the investment, $\Theta$ is a risk-free return rate, such as the return rate of bank.
\end{Def}
Given a sequential investment with $T$ holding periods, $A_T$ is calculated as
\begin{equation}\label{}\small
    A_T = \frac{1}{T} \sum_{t=1}^T R_t - TC_t,
\end{equation}
where $TC_t$ is a transaction cost in the $t$-th period. The volatility $V_T$ in Eq.~\eqref{eq:sharpe_ratio} is defined as
\begin{equation}\label{}\small
  V_T = \sqrt{\frac{\sum_{t=1}^T (R_t - \bar{R}_t)^2 }{T}},
\end{equation}
where $\bar{R}_t = \sum_{t=1}^T R_t/T$ is the average of $R_t$.

For a $T$-period investment, the optimization objective of our strategy is to generate the long and short portfolio sequences $\bm{B}^+ =  \{\bm{b}^+_{1}, \ldots \bm{b}^+_{T}\}$ and $\bm{B}^- =  \{\bm{b}^-_{1}, \ldots, \bm{b}^-_{T}\}$ that can maximize the Sharpe ratio of the investment as
\begin{equation}\label{eq:obj}\small
    \mathop{\arg\max}_{\{\bm{B}^+,\bm{B}^-\}} H_T\left(\bm{B}^+,\bm{B}^-\right).
\end{equation}

\paratitle{Insight II.}  The Sharpe ratio evaluates the performance of a strategy from both profit and risk perspectives. This profit-risk balance characteristic requires our model not only focuses on maximizing return rate $R_t$ for each period, but also considers the long-term volatility of $R_t$ across all periods in an investment. In other words, designing a far-sighted steady investment strategy is more valuable than a short-sighted strategy with short-term high profits.

\section{The \name Model}


\begin{figure}[t]
\includegraphics[width=0.85\columnwidth]{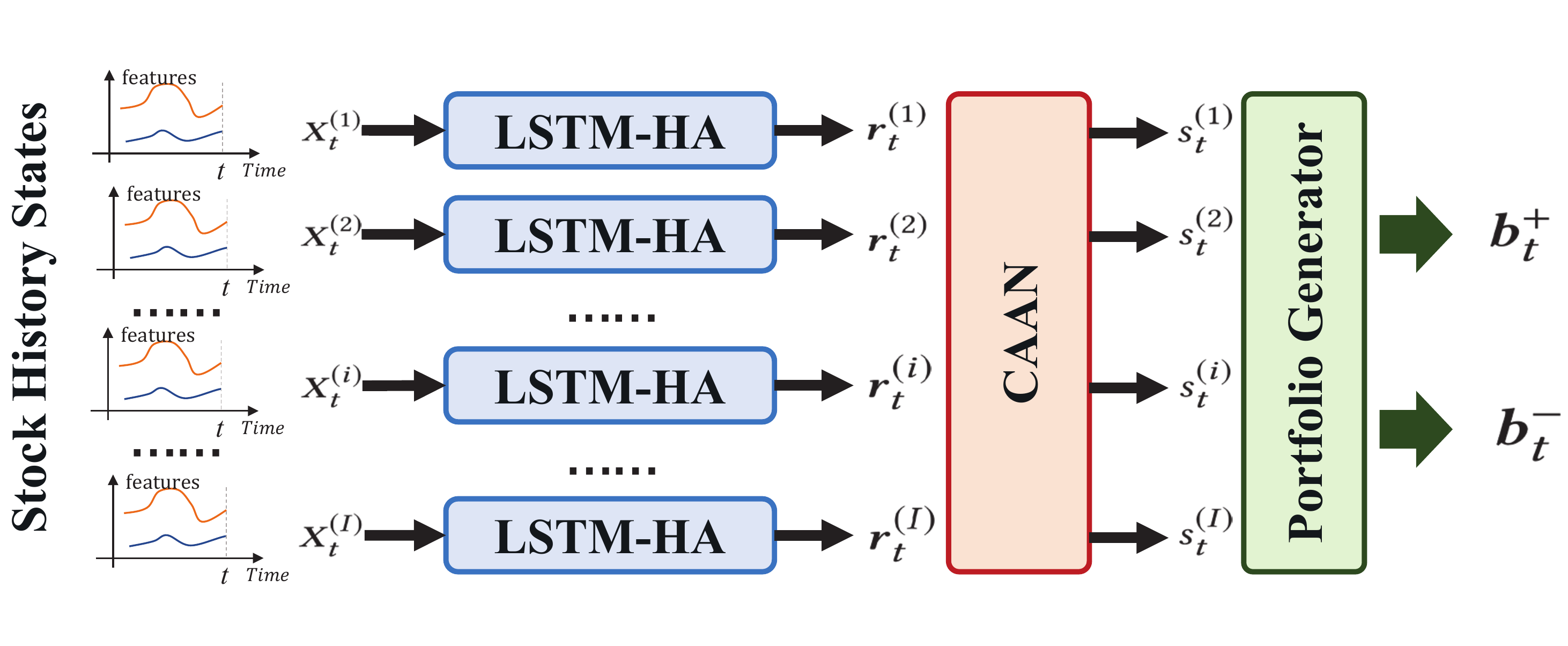}
\caption{The framework of the \name model.}\label{fig:framework}\vspace{-0.5cm}
\end{figure}


In this section, we propose a reinforcement learning (RL) based model called \emph{\name} to implement a BWSL strategy with the Sharpe ratio defined in Eq.~\eqref{eq:sharpe_ratio} as the optimization objective. As shown in Fig.~\ref{fig:framework}, \name contains three components. The first component is a LSTM with History state Attention network (LSTM-HA). For each stock $i$, we use the LSTM-HA model to extract a stock representation $\bm{r}^{(i)}$ from its history states $\bm{X}^{(i)}$. The second component is a Cross-Asset Attention Network (CAAN) to describe interrelationships among the stocks. The CAAN takes as input the representations ($\bm{r}^{(i)}$) of all stocks, and estimates a winner score $s^{(i)}$ for every stock. The $s^{(i)}$ is a score to indicate the degree of stock $i$ belonging to a winner. The third component is a portfolio generator, which calculates the investment proportions in $\bm{b}^{+}$ and $\bm{b}^{-}$ according to the scores ($s^{(i)}$) of all stocks. We use reinforcement learning to end-to-end optimize the three components as a whole, where the Sharpe ratio of a sequential investment is maximized through a far-sighted way.

\subsection{Raw Stock Features}~\label{sec:features}

The stock features used in our model contains two categories. The first category is the {\em trading features}, which describes the trading information of a stock. At time $t$, the trading features include:

{$\bullet$ \bf Price Rising Rate (PR)}: The price rising rate of a stock during the last holding period. It is defined as $\left({p_{t}^{(i)}}/{p_{t-1}^{(i)}}\right)$ for stock $i$.

{$\bullet$ \bf Fine-grained Volatility (VOL)}: A holding period can be further divided into many sub-periods. We set one month as a holding period in our experiment, thus a sub-period can be a trading day. VOL is defined as the standard deviation of the prices of all sub-periods from $t-1$ to $t$.

{$\bullet$ \bf Trade Volume (TV)}: The total quantity of stocks traded from $t-1$ to $t$. It reflects the market activity of a stock.

The second category is the company features, which describe the financial condition of the company that issues a stock. At time $t$, the company features include:

{$\bullet$ \bf Market Capitalization (MC)}: For stock $i$, it is defined as the product of the price $p_t^{(i)}$ and the outstanding shares of the stock.

{$\bullet$ \bf Price-earnings Ratio (PE)}: It is the ratio of the market capitalization of a company to its annual earnings.

{$\bullet$ \bf Book-to-market Ratio (BM)}: It is the ratio of the book value of a company to its market value.

{$\bullet$ \bf Dividend (Div)}: It is the reward from company's earnings to stock holders during the $(t-1)$-th holding period.

Since the values of these features are not in the same scale, we standardize them into Z-scores. 

\subsection{Stock Representations Extraction}~\label{sec:extracting}

The performance of a stock has close relations with its history states. In the \name model, we propose a {\em Long Short-Term Memory with History state Attention (LSTM-HA)} model to learn the representation of a stock from its history features. 


\paratitle{The sequential representation.}
In the LSTM-HA network, we use the vector $\tilde{\bm{x}}_t$ to denote the history state of a stock at time $t$, which consists of the stock features given in Section~\ref{sec:features}. We name the last $K$ historical holding periods at time $t$, \ie the period from time $t-K$ to time $t$, as a {\em look-back window} of $t$. The history states of a stock in the look-back window are denoted as a sequence $\bm{X} = \{\bm{x}_{1},$ $\ldots,$ $\bm{x}_{k},$ $\ldots,$ $\bm{x}_{K}\}$~\footnote{We also use $\bm{X}$ to denote the matrix $\left(\bm{x}_{k}\right)$, the two definitions are interchangeable.}, where $\bm{x}_{k} = \tilde{\bm{x}}_{t-K+k}$. Our model uses a Long Short-Term Memory (LSTM) network~\cite{hochreiter1997long} to recursively encode $\bm{X}$ into a vector as
\begin{equation}\label{eq:lstm}\small
    \bm{h}_k = \mathrm{LSTM}\left(\bm{h}_{k-1}, \bm{x}_{k} \right),~k\in[1,K]
\end{equation}
where $\bm{h}_k$ is the hidden state encoded by LSTM at step $k$. The $\bm{h}_K$ at the last step is used as a representation of the stock. It contains the sequential dependence among elements in $\bm{X}$.

\paratitle{The history state attention.}
The $\bm{h}_K$ can fully exploit the sequential dependence of elements in $\bm{X}$, but the global and long-range dependence among $\bm{X}$ are not effectively modeled. Therefore, we adopt a history state attention to enhance $\bm{h}_K$ using all middle hidden states $\bm{h}_k$. Specifically, following the standard attention~\cite{sutskever2014sequence}, the history state attention enhanced representation, denoted as $\bm{r}$, is calculated as
\begin{equation}\label{eq:lstm-ha_3}
  \bm{r} = \sum_{k=1}^{K} \mathrm{ATT}\left(\bm{h}_K, \bm{h}_k\right) \bm{h}_k,
\end{equation}
where $\mathrm{ATT}(\cdot,\cdot)$ is an attention function defined as
\begin{eqnarray}\label{eq-att}\small
    \mathrm{ATT}\left(\bm{h}_K, \bm{h}_k\right)&=& \frac{\exp\left(\alpha_{k}\right)}{\sum_{k'=1}^{K}\exp\left(\alpha_{k'}\right)},  \\\nonumber
    \alpha_k &=& \bm{w}^{\top} \cdot \tanh \left(\bm{W}^{(1)}\bm{h}_k + \bm{W}^{(2)}\bm{h}_K \right).\nonumber
\end{eqnarray}
Here, $\bm{w}$, $\bm{W}^{(1)}$ and $\bm{W}^{(2)}$ are the parameters to learn.

For the $i$-th stock at time $t$, the history state attention enhanced representation is denoted as $\bm{r}_t^{(i)}$. It contains both the sequential and global dependences of stock $i$'s history states from time $t-K+1$ to time $t$. In our model, the representation vectors for all stocks are extracted by the same LSTM-HA network. The parameters $\bm{w}$, $\bm{W}^{(1)}$, $\bm{W}^{(2)}$ and those of the LSTM network in Eq.~\eqref{eq:lstm} are shared by all stocks. In this way, the representations extracted by LSTM-HA are relatively stable and general for all stocks rather than for a particular one.

\paratitle{Remark.} A major advantage of LSTM-HA is that it can learn both the sequential and global dependences from stock history states. Compared with the existing studies that only use a recurrent neural network to extract the sequential dependence in history states~\cite{moody1998performance,deng2017deep} or directly stack history states as an input vector of MLP~\cite{dql_stanford} to learn the global dependence, our model describes stock histories more comprehensively. It is worth mentioning that LSTM-HA is also an open framework. The representations learned from other types of information sources, such as news, events and social media~\cite{news1,event1,tweets}, could also be concatenated or attended with $\bm{r}^{(i)}_t$.

\subsection{Winners and Losers Selection}

In the traditional RL-based strategy models, the investment portfolio is often directly generated from the stock representations through a softmax normalization~\cite{dql_stanford,deng2017deep,ding2018investor}. The drawback of this type of methods is that it does not fully exploit the interrelationships among stocks, which however is very important for the BWSL strategy as analyzed in \emph{Insight I} of Section~\ref{sec:define}. In light of this, we propose a {\em Cross-Asset Attention Network} (CAAN) to describe the interrelationships among stocks.


\paratitle{The basic CAAN model.}
The CAAN model adopts the self-attention mechanism proposed by Ref.~\cite{transformer} to model the interrelationships among stocks. Specifically, given the stock representation $\bm{r}^{(i)}$ (we omit time $t$ without loss of generality), we calculate a query vector $\bm{q}^{(i)}$, a key vector $\bm{k}^{(i)}$ and a value vector $\bm{v}^{(i)}$ for stock $i$ as
\begin{equation}\label{eq:qkv}\small
  \bm{q}^{(i)} = \bm{W}^{(Q)} \bm{r}^{(i)},~~~
  \bm{k}^{(i)} = \bm{W}^{(K)} \bm{r}^{(i)},~~~
  \bm{v}^{(i)} = \bm{W}^{(V)} \bm{r}^{(i)},
\end{equation}
where $\bm{W}^{(Q)}$, $\bm{W}^{(K)}$ and $\bm{W}^{(V)}$ are the parameters to learn. The interrelationship of stock $j$ to stock $i$ is modeled as using the $\bm{q}^{(i)}$ of the stock $i$ to query the key $\bm{k}^{(j)}$ of stock $j$, \ie
\begin{equation}\label{eq:self-attention1}\small
  \beta_{ij} = \frac{\bm{q}^{(i)\top}\cdot \bm{k}^{(j)}}{\sqrt{D_k}},
\end{equation}
where $D_k$ is a re-scale parameter setting following Ref.~\cite{transformer}. Then, we use the normalized interrelationships $\{\beta_{ij}\}$ as weights to sum the values $\{\bm{v}^{(j)}\}$ of other stocks into an attenuation score:
\begin{equation}\label{eq:self-attention3}\small
  \bm{{a}}^{(i)} = \sum_{j=1}^I \mathrm{SATT}\left(\bm{q}^{(i)}, \bm{k}^{(j)}\right) \cdot \bm{v}^{(j)},
\end{equation}
where the self-attention function $\mathrm{SATT}\left(\cdot, \cdot\right)$ is a softmax normalized interrelationships of $\beta_{ij}$, \ie
\begin{equation}\label{eq:self-attention2}\small
  \mathrm{SATT}\left(\bm{q}^{(i)}, \bm{k}^{(j)}\right) = \frac{\exp\left(\beta_{ij}\right)}{\sum_{j'=1}^{I}\exp\left(\beta_{ij'}\right)}.
\end{equation}
We use a fully connected layer to transform the attention vector $\bm{{a}}^{(i)}$ into a winner score as
\begin{equation}\label{eq:winner_score}\small
  s^{(i)} = \mathrm{sigmoid}\left(\bm{w}^{(s)\top}\cdot \bm{{a}}^{(i)} + e^{(s)}\right),
\end{equation}
where $\bm{w}^{(s)}$ and $e^{(s)}$ are the connection weights and the bias to learn. The winner score $s^{(i)}_t$ indicates the degree of stock $i$ being a winner in the $t$-th holding period. A stock with a higher score is more likely to be a winner.

\paratitle{Incorporating price rising rank prior.}
In the basic CAAN, the interrelationships modeled by Eq.~\eqref{eq:self-attention1} are directly learned from data. In fact, we could use priori knowledge to help our model to learn the stock interrelationships. We use $c_{t-1}^{(i)}$ to denote the rank of price rising rate of stock $i$ in the last holding period (from $t-1$ to $t$). Inspired by the method for modeling positional information from the NLP field, we use the relative positions of stocks in the coordinate axis of $c_{t-1}^{(i)}$ as a priori knowledge of the stock interrelationships. Specifically, given two stocks $i$ and $j$, we calculate their discrete relative distance in the coordinate axis of $c_{t-1}^{(i)}$ as
\begin{equation}\label{eq:relative_distance}\small
  d_{ij} = \left\lfloor{\left|c_{t-1}^{(i)} - c_{t-1}^{(j)}\right|}\Big/{Q}\right\rfloor,
\end{equation}
where $Q$ is a preset quantization coefficient. We use a lookup matrix $\bm{L} = (\bm{l}_1, \ldots, \bm{l}_L)$ to represent each discretized value of $d_{ij}$. Using the $d_{ij}$ as the index, the corresponding column vector $\bm{l}_{d_{ij}}$ is an embedding vector of the relative distance $d_{ij}$. 

For a pair of stocks $i$ and $j$, we calculate a priori relation coefficient $\psi_{ij}$ using $\bm{l}_{d_{ij}}$ as
\begin{equation}\label{}\small
    \psi_{ij} = \mathrm{sigmoid} \left(\bm{w}^{(L)\top}\bm{l}_{d_{ij}} \right),
\end{equation}
where $\bm{w}^{(L)}$ is a learnable parameter. The relationship between $i$ and $j$ estimated by Eq.~\eqref{eq:self-attention1} is rewritten as
\begin{equation}\label{eq:winner_score2}\small
   \beta_{ij} = \frac{\psi_{ij}\left(\bm{q}^{(i)\top}\cdot \bm{k}^{(j)}\right)}{\sqrt{D}}.
\end{equation}
In this way, the relative positions of stocks in price rising rate rank are introduced as a weight to enhance or weaken the attention coefficient. The stocks have similar history price rising rates will have a stronger interrelationship in the attention and then have similar winner scores. 

\paratitle{Remark.} As shown in Eq.~\eqref{eq:self-attention3}, for each stock $i$, the winner score $s^{(i)}$ is calculated according to the attention of all other stocks. In this way, the interrelationships among all stocks are involved into CAAN. This special attention mechanism meets the model design requirement of {\em Insight I} in Section~\ref{sec:define}.

\subsection{Portfolios Generator}\label{sec:portfolio}

Given the winner scores $\{s^{(1)},$ $\ldots,$ $s^{(i)},$ $\ldots,$ $s^{(I)} \}$ of $I$ stocks, our \name model generally buys the stocks with high winner scores and sells those with low winner scores. Specifically, we first sort the stocks in descending order by their winner scores and obtain the sequence number $o^{(i)}$ for each stock $i$. Let $G$ denote the preset size of portfolio $\bm{b}^+$ and $\bm{b}^-$. If $o^{(i)} \in [1,G]$, stock $i$ will enter the portfolio $\bm{b}^{+(i)}$, with the investment proportion calculated as
\begin{equation}\label{eq:long_portfolio_g}\small
  b^{+(i)} = \frac{\exp\left(s^{(i)}\right)}{\sum_{o^{(i')} \in [1,G]} \exp\left(s^{(i')}\right)}.
\end{equation}
If $o^{(i)} \in (I-G,I]$, stock $i$ will enter $\bm{b}^{-(i)}$ with a proportion
\begin{equation}\label{eq:short_portfolio_g}\small
  b^{-(i)} = \frac{\exp\left(1-s^{(i)}\right)}{\sum_{o^{(i')} \in (I-G,I]} \exp\left(1-s^{(i')}\right)}.
\end{equation}
The rest stocks are unselected for the lack of clear buy/sell signals. For simplicity, we can use one vector to record all the information of the two portfolios. That is, we form the vector $\bm{b}^c$ of length $I$, with $b^{c(i)}=b^{+(i)}$ if $o^{(i)} \in [1,G]$, or $b^{c(i)}=b^{-(i)}$ if $o^{(i)} \in (I-G,I]$, or 0 otherwise, $i=1,\ldots,I$. In what follows, we use $\bm{b}^c$ and $\{\bm{b}^{+},\bm{b}^{-}\}$ interchangeably as the return of our \name model for clarity.


\subsection{Optimization via Reinforcement Learning}

{
We frame the \name strategy into a RL game with discrete agent actions to optimize the model parameters, where a $T$-period investment is modeled as a state-action-reward trajectory $\pi$ of a RL agent, \ie $\pi = \{state_1,$ $action_1,$ $reward_1,$ $\ldots,$ $state_t,$ $action_t,$ $reward_t,$ $\ldots,$ $state_T,$ $action_T,$ $reward_T\}$. The $state_t$ is the history market state observed at $t$, which is expressed as $\mathcal{{X}}_t = (\bm{X}_t^{(i)})$. The $action_t$ is an $I$-dimensional binary vector, of which the element $action_t^{(i)} = 1$ when the agent invests stock $i$ at $t$, and $0$ otherwise\footnote{In the RL game, the actions of an agent are discrete states with the probability ${b}_t^{c(i)}/2$ indicating whether to invest stock $i$. In the real investment, we allocate capitals to stocks $i$ according the continuous proportion ${b}_t^{c(i)}$. This approximation is for the sake of problem solving.}. According to $state_t$, the agent has a probability $\mathrm{Pr}(action_t^{(i)} = 1)$ to invest stock $i$, which is determined by \name as
\begin{equation}\label{}\small
  \mathrm{Pr}\left(action_t^{(i)} = 1 \big |\mathcal{X}^{n}_{t}, \theta\right) = \frac{1}{2}\mathcal{G}^{(i)}(\mathcal{X}^{n}_{t}, \theta) = \frac{1}{2}{b}_t^{c(i)},
\end{equation}
where $\mathcal{G}^{(i)}(\mathcal{X}^{n}_{t}, \theta)$ is part of \name that generates ${b}_t^{c(i)}$, $\theta$ denotes the model parameters, and $1/2$ is to ensure $\sum_{i=1}^{I} \mathrm{Pr}(action_t^{(i)} = 1) = 1$. Let ${H}_\pi$ denote the Sharpe ratio of $\pi$, then $reward_t$ is the contribution of $action_t$ to ${H}_\pi$, with $\sum_{t=1}^{T} reward_t = {H}_\pi$.

For all possible $\pi$, the average reward of the RL agent is
\begin{equation}\label{eq:average_reward}\small
  J(\theta) = \int_\pi {H}_\pi \mathrm{Pr}(\pi|\theta) \mathrm{d}\pi,
\end{equation}
where $\mathrm{Pr}(\pi|\theta)$ is the probability of generating $\pi$ from $\theta$. Then, the objective of the RL model optimization is to find the optimal parameters $\theta^* =\mathop{\arg\max}_{\theta} J(\theta)$.

We use the gradient ascent approach to iteratively optimize $\theta$ at round $\tau$ as $\theta_\tau = \theta_{\tau-1} + \left.\eta \nabla J(\theta)\right|_{\theta = \theta_{\tau-1}}$, where $\eta$ is a learning rate. Given a training dataset that contains $N$ trajectories $\{\pi_1,$ $\ldots,$ $\pi_n,$ $\ldots,$ $\pi_N\}$, $\nabla J(\theta)$ can be approximately calculated as~\cite{RL}
\begin{equation}\label{eq:derivative}\small
\begin{aligned}
  \nabla J(\theta)  = &\int_\pi {H}_\pi \mathrm{Pr}(\pi|\theta) \nabla \log \mathrm{Pr}(\pi|\theta) \mathrm{d}\pi.\\
    \approx  &\frac{1}{N} \sum_{n=1}^N \left( {H}_{\pi_n} \sum_{t=1}^{T_n} \sum_{i=1}^{I} \nabla_\theta \log \mathrm{Pr}\left(action_t^{(i)} = 1 \big|\mathcal{X}^{(n)}_{t}, \theta\right)\right),
\end{aligned}
\end{equation}
The gradient $\nabla_\theta \log \mathrm{Pr}(action_t^{(i)} = 1 |\mathcal{X}^{(n)}_{t}, \theta)=\nabla_\theta \log \mathcal{G}^{(i)}(\mathcal{X}^{n}_{t}, \theta)$, which is calculated by the Back Propagation algorithm.
}

In order to ensure the proposed model can beat the market, we introduce the threshold method~\cite{RL} into our reinforcement learning. Then the gradient $\nabla J(\theta)$ in Eq.~\eqref{eq:derivative} is rewritten as
\begin{equation}\label{eq:derivative2}\small
  \nabla J(\theta) = \frac{1}{N} \sum_{n=1}^N  \left(\left({H}_{\pi_n} - H_0\right) \sum_{t=1}^{T_n}\sum_{i=1}^{I} \nabla_\theta \log \mathcal{G}^{(i)}\left(\mathcal{X}^{n}_{t}, \theta\right)\right),
\end{equation}
where the threshold $H_0$ is set as the Sharpe ratio of the overall market. In this way, the gradient ascent only encourages the parameters that can outperform the market.

{\bf Remark.} The Eq.~\eqref{eq:derivative2} uses $({H}_{\pi_n} - H_0)$ to integrally weight the the gradients  $\nabla_\theta \log \mathcal{G}$ of all holding periods in $\pi_n$. The reward is not directly given to any isolated step in $\pi_n$ but given to all steps in $\pi_n$. This feature of our model meets the far-sight requirement of {\em Insight II} in Section~\ref{sec:define}.

\section{Model Interpretation}
\label{sec:interpret}

In the \name model, the LSTM-HA and CAAN networks cast the raw stock features as winner scores. The final investment portfolios are directly generated from the winner scores. A natural follow-up question is: what kind of stocks would be selected as winners by \name? To answer this question, we propose a sensitivity analysis method~\cite{sensitivity,wavelet,icdm} to interpret how the history features of a stock influence its winner score in our model.

We use $s = \mathcal{F}(\bm{X})$ to express the function of history features $\bm{X}$ of a stock to its winner score $s$. In our model, $s = \mathcal{F}(\bm{X})$ is a combined network of LSTM-HA and CAAN. We use $x_q$ to denote an element of $\bm{X}$ which is the value of one feature (defined in Section~\ref{sec:features}) at a particular time period of the look-back window, \eg the price rising rate of a stock at the time of three months ago.

Given the history state $\bm{X}$ of a stock, the influence of $x_q$ to its winner score $s$, \ie the sensitivity of $s$ to $x_q$, is expressed as
\begin{equation} \label{eq:df_da}\small
\begin{aligned}
  \delta_{x_q}\left(\bm{X}\right) & = \lim_{\Delta x_q \rightarrow 0} \frac{\mathcal{F}\left({\bm{X}}\right) - \mathcal{F}\left(x_q + \Delta x_q, {\bm{X}}_{\neg x_q}\right)}{x_q - \left(x_q + \Delta x_q\right)} = \frac{\partial \mathcal{F}\left({\bm{X}}\right)}{\partial x_q},
\end{aligned}
\end{equation}
where ${\bm{X}}_{\neg x_q}$ denotes the elements of ${\bm{X}}$ except $x_q$. 


For all possible stock states in a market, the average influence of the stock state feature $x_q$ to the winner score $s$ is
\begin{equation}\label{eq:int_a}\small
  \bar{\delta}_{x_q} = \int_{D_{\bm{X}}} \mathrm{Pr}(\bm{X}) \delta_{x_q}(\bm{X})~\mathrm{d}_\sigma.
\end{equation}
%
where $\mathrm{Pr}(\bm{X})$ is the probability density function of $\bm{X}$, and $\int_{D_{\bm{X}}} \cdot~\mathrm{d}_\sigma$ is an integral over all possible value of ${\bm{X}}$. According to the Large Number Law, given a dataset that contains history states of $I$ stocks in $N$ holding periods, the $\bar{\delta}_{x_q}$ is approximated as
\begin{equation}\label{eq:delta_bar}\small
  \bar{\delta}_{x_q} = \frac{1}{I\times N}\sum_{n=1}^N \sum_{i=1}^I {\delta}_{x_q}\left(\bm{X}^{(i)}_n \Big| \mathcal{X}^{(\neg i)}_n\right),
\end{equation}
where $\bm{X}^{(i)}_n$ is the history state of the $i$-th stock at the $n$-th holding period, and $\mathcal{X}^{(\neg i)}_n$ denotes the history states of other stocks that are concurrent with the history state of $i$-th stock.


We use $\bar{\delta}_{x_q}$ to measure the overall influence of a stock feature $x_q$ to the winner score. A positive value of $\bar{\delta}_{x_q}$ indicates that our model tends to take a stock as a winner when $x_q$ is large, and vice versa. For example, in the experiment to follow, we obtain $\bar{\delta}<0$ for the fine-grained volatility feature, which means that our model trends to select low volatility stocks as winners.


\section{Experiment}

In this section, we empirically evaluate our \name model by the data in the U.S. markets.  The data in the Chinese stock markets are also used for robustness check. 

\subsection{Data and Experimental Setup}\label{sec:setup}

The data of U.S. stock market used in our experiments are obtained from Wharton Research Data Services (WRDS)~\footnote{\url{https://wrds-web.wharton.upenn.edu/wrds/}}. The time range of the data is from Jan. 1970 to Dec. 2016. This long time range covers several well-known market events, such as the dot-com bubble from 1995 to 2000 and the subprime mortgage crisis from 2007 to 2009, which enables the evaluation over diverse market states. The stocks are from four markets: NYSE, NYSE American, NASDAQ, and NYSE Arca. The number of valid stocks is more than 1000 per year. We use the data from Jan. 1970 to Jan. 1990 as the training and validation set, and the rest as the test set. 

In the experiment, the holding period is set to one month, and the number of holding periods $T$ in an investment is set to 12, \ie the Sharpe ratio reward is calculated every 12 months for RL. The look-back window size $K$ is set to 12, \ie we look back on the 12-month history states of stocks. The size $G$ of the portfolios is set as 1/4 of number of all stocks. 




\subsection{Baseline Methods}

\name is compared with a number of baselines including:

$\bullet$ \emph{Market}: the uniform Buy-And-Hold strategy~\cite{huang2016robust};

$\bullet$ Cross Sectional Momentum (\emph{CSM})~\cite{jegadeesh2002cross} and Time Series Momentum (\emph{TSM})~\cite{moskowitz2012time}: two classic momentum strategies;

$\bullet$ Robust Median Reversion (\emph{RMR}): a newly reported reversion strategy~\cite{huang2016robust};

$\bullet$ Fuzzy Deep Direct Reinforcement (\emph{FDDR}): a newly reported RL-based BWSL strategy~\cite{deng2017deep};

$\bullet$ \name-NC (\emph{AS-NC}): the \name model without the CAAN, where the outputs of LSTM-HA are directly used as the inputs of the portfolio generator.

$\bullet$ \name-NP (\emph{AS-NP}): the \name model without price rising rank prior, where we use the basic CAAN in our model.

The baselines TSM/CSM/RMR represent the traditional financial strategies. TSM and CSM are based on the momentum logic and RMR is based on the reversion logic. FDDR represents the state-of-the-art RL-based BWSL strategy. AS-NC and AS-NP are used as a contrast to verify the effectiveness of the CAAN and  price rising rank prior. The Market is used to indicate states of the market.

\subsection{Evaluation Measures}

The most standard evaluation measure for investment strategies is {\em Cumulative Wealth}, which is defined as
\begin{equation}\label{}
  CW_T = \prod_{t=1}^{T} \left(R_t + 1 - TC\right),
\end{equation}
where $R_t$ is the rate of return defined in Eq.~\eqref{eq:return_rate} and the transaction cost $TC$ is set to 0.1\% in our experiments according to Ref.~\cite{deng2017deep}.

The preferences of different investors are varied. Therefore, we also use some other evaluation measures including:

1) {\em Annualized Percentage Rate (APR)} is an annualized average of return rate. It is defined as $APR_T = A_T \times N_Y$, where $N_Y$ is the number of holding periods in a year. 

2) {\em Annualized Volatility (AVOL)} is an annualized average of volatility. It is defined as $AVOL_T = {V_T}\times \sqrt{N_Y}$ and is used to measure the average risk of a strategy during an unit time period.

3) {\em Annualized Sharpe Ratio (ASR)} is the risk-adjusted annualized return based on APR and AVOL. The formalized definition of ASR is $ASR_T = {APR_T}/{AVOL_T}$.

4) {\em Maximum DrawDown (MDD)} is the maximum loss from a peak to a trough of a portfolio, before a new peak is attained. It is the other way to measure the investment risk. The formalized definition of MDD is
\begin{equation}\label{}\small
MDD_T = \max_{\tau \in [1,T]} \left(\max_{t \in [1,\tau]} \left(\frac{APR_t - APR_{\tau}}{APR_t}\right) \right).
\end{equation}

5) {\em Calmar Ratio (CR)} is the risk-adjusted APR based on Maximum DrawDown. It is calculated as $CR_T = {APR_T}/{MDD_T}$.

6) {\em Downside Deviation Ratio (DDR)} measures the downside risk of a strategy as the average of returns when it falls below a minimum acceptable return (MAR). It is the risk-adjusted APR based on Downside Deviation. The formalized definition of DDR is given as
\begin{equation}\label{}\small
DDR_T = \frac{APR_T}{\textup{Downside Deviation}} = \frac{APR_T}{\sqrt{\mathbb{E}[\min(R_t,MAR)]^2}},~~t\in [1,T].
\end{equation}
In our experiment, the MAR is set to zero.

\subsection{Performance in U.S. Markets}
\begin{figure}[t]
	\includegraphics[width=0.75\columnwidth]{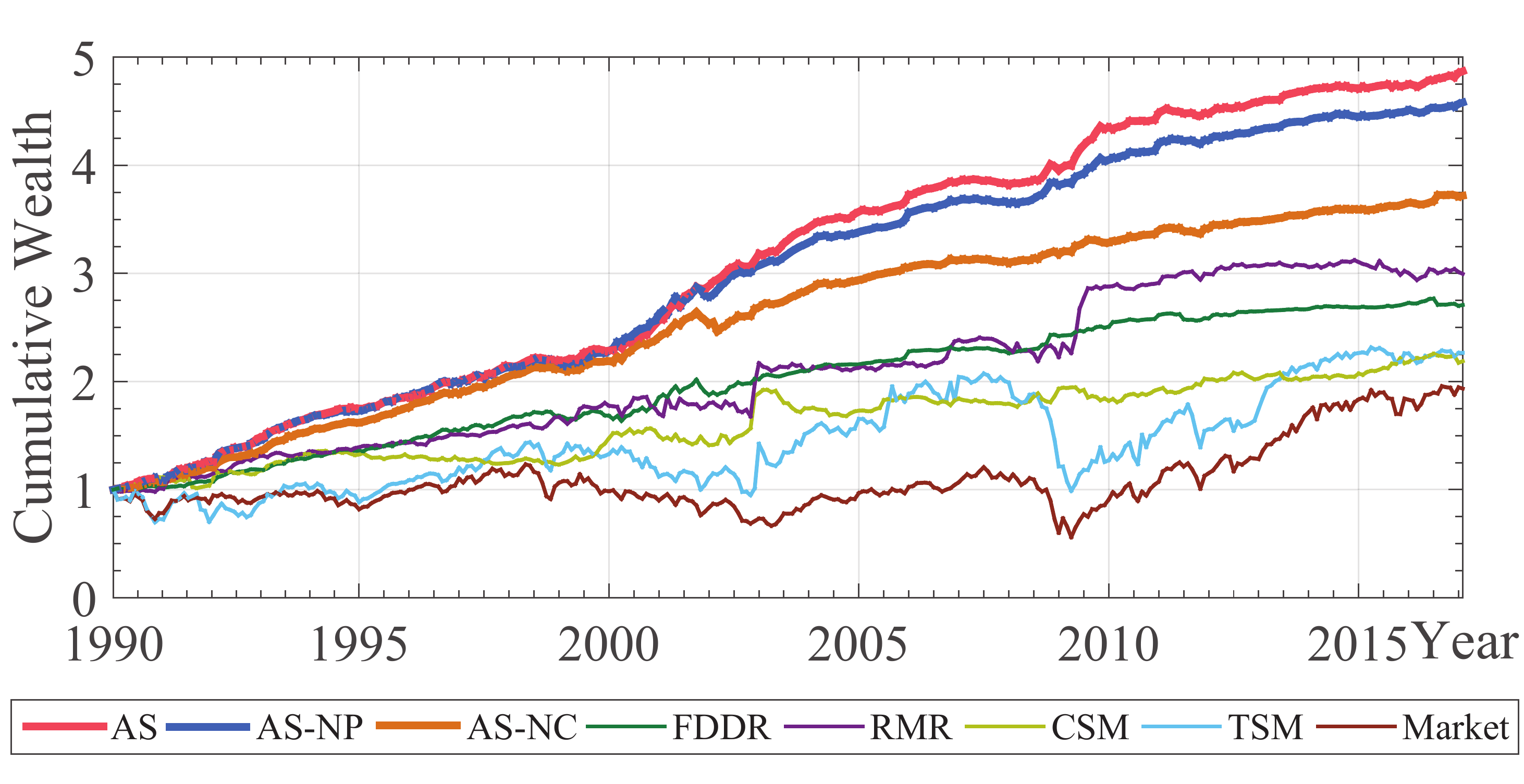}
	\vspace{-0.3cm}
	\caption{The Cumulative Wealth in U.S. markets.}\vspace{-0.3cm}
	\label{fig:cw}
\end{figure}

Fig.~\ref{fig:cw} is a cumulative wealth comparison of \name and the baselines. In general, the performance of \name (AS) is much better than other baselines, which verifies the effectiveness of our model. Some interesting observations are highlighted as follows:

1) The performance of \name is better than \name-NP and the performance of \name-NP is better than \name-NC, which indicates that the stock rank priors and interrelationships modeled by CAAN are very helpful for the BWSL strategy.

2) The FDDR is also a kind of deep RL investment strategy, which extracts the fuzzy representations of stocks using a recurrent deep neural network. In our experiment, the performance of \name-NC is better than FDDR, indicating the advantage of our LSTM-HA network in the stock representation learning.

3) The TSM strategy performs well in the bull market but very poorly in the bear market (the financial crisis in 2003 and 2008), while the RMR has an opposite performance. This implies the traditional financial strategies can only adapt to a certain type of market state without an effective forward-looking mechanism. This defect is greatly addressed by the RL strategies, including \name and FDDR, which perform much stably across different market states. 

\begin{table}[t]\footnotesize
  \caption{Performance comparison on U.S. markets.}
  \label{tab:metrics}\vspace{-0.3cm}
  \begin{tabular}{lcccccc}
    \toprule
    &{\bf APR}&{\bf \underline{AVOL}}&{\bf ASR}&{\bf \underline{MDD}}&{\bf CR}&{\bf DDR}\\
    \midrule
    {\bf Market} & 0.042 & 0.174 & 0.239 & 0.569 & 0.073 & 0.337 \\
    {\bf TSM} & 0.047 & 0.223 & 0.210 &0.523 & 0.090 & 0.318 \\
    {\bf CSM} & 0.044 & 0.096 & 0.456 & 0.126 & 0.350 & 0.453 \\
    {\bf RMR} & 0.074 & 0.134 & 0.551 & 0.098 & 1.249 & 0.757 \\
    {\bf FDDR}& 0.063 & 0.056 & 1.141 & 0.070 & 0.900 & 2.028\\
    \midrule
    {\bf AS-NC} & 0.101 & {\bf 0.052} & 1.929 & 0.068 & 1.492 & 1.685 \\
    {\bf AS-NP} & 0.133 & 0.065 &  2.054  &  0.033 & 3.990 & 4.618  \\
    {\bf AS} & {\bf 0.143} & 0.067 & {\bf 2.132}  & {\bf 0.027} & {\bf 5.296} & {\bf 6.397}  \\
  \bottomrule
\end{tabular}\vspace{-0.3cm}
\end{table}

The performances evaluated by other measures are listed in Table~\ref{tab:metrics}. For the measures underlined (AVOL, MDD), the lower value indicates the better performance, while the situation is opposite for the other measures. As shown in Table~\ref{tab:metrics}, the performances of \name, \name-NP and \name-NC are better than other baselines with all measures, confirming the effectiveness and robustness of our strategy. The performances of \name, \name-NP and \name-NC are close in terms of ASR, which might be due to all of these models are optimized for maximizing the Sharpe ratio. The profits of \name and \name-NP measured by APR are higher than that of \name-NC, at the cost of a little bit higher volatility.

More interestingly, the performance of \name measured by MDD, CR and DDR is much better than that of \name-NP. The similar results could be observed by comparing MDD, CR and DDR of \name-NP and \name-NC. The three measures are used to indicate the extreme loss in an investment, \ie the maximum draw down and the returns below the minimum acceptable threshold. The results suggest that the extreme loss control ability of the three models are \name $>$ \name-NP $>$ \name-NC, which highlights the contribution of the CAAN component and the price rising rank prior. Indeed, CAAN with price rising rank priors fully exploits the ranking relationship among stocks. This mechanism can protect our strategy from the error of ``buying losers and selling winners'', and therefore can greatly avoid extreme losses in investments. In summary, \name is a very competitive strategy for investors with different types of preferences.


\begin{figure*}[t]
\subfigure[Price Rising]{\includegraphics[width=0.50\columnwidth]{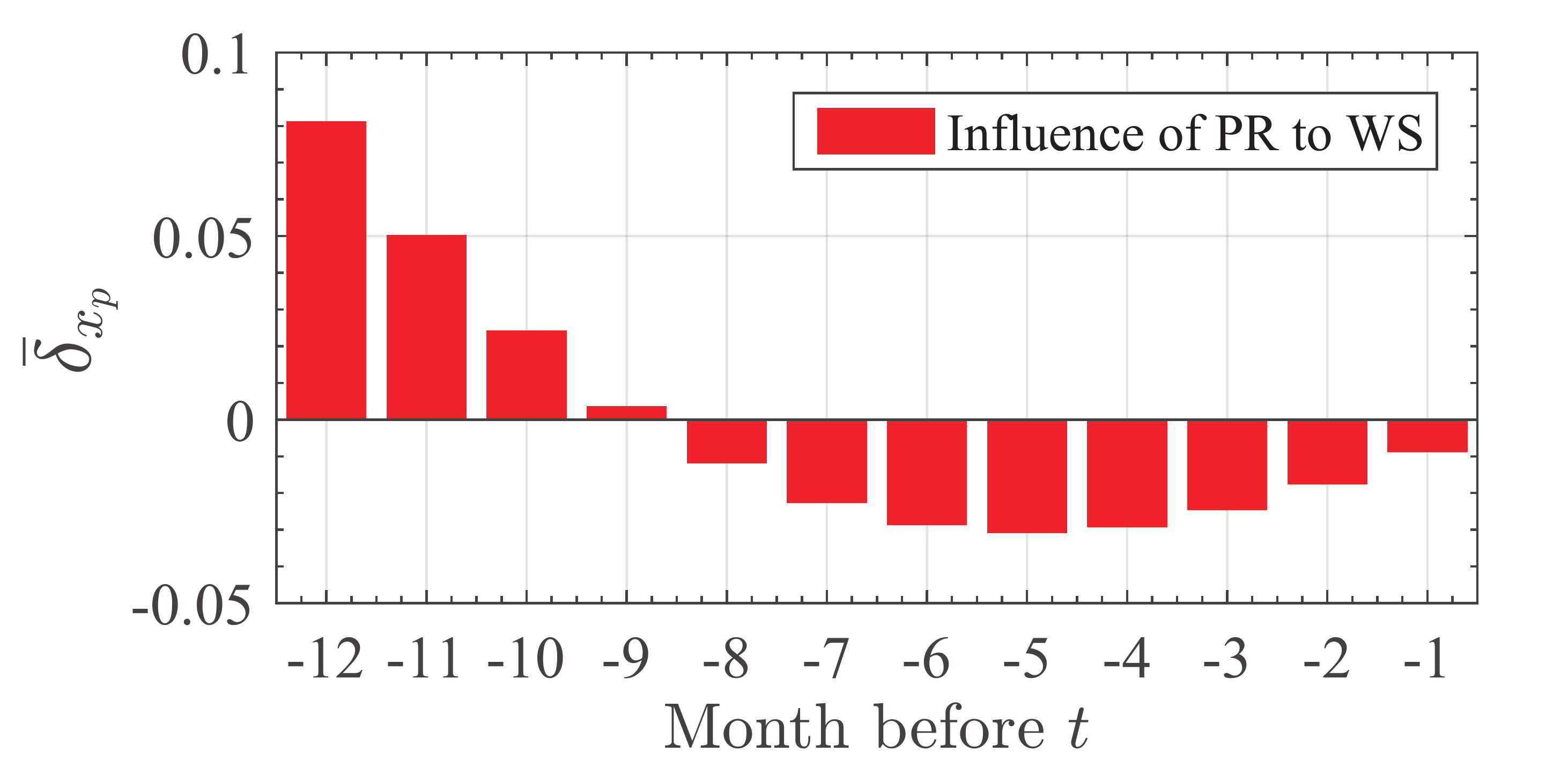}\label{fig:PR_influence}}
\subfigure[Trade Volume]{\includegraphics[width=0.5\columnwidth]{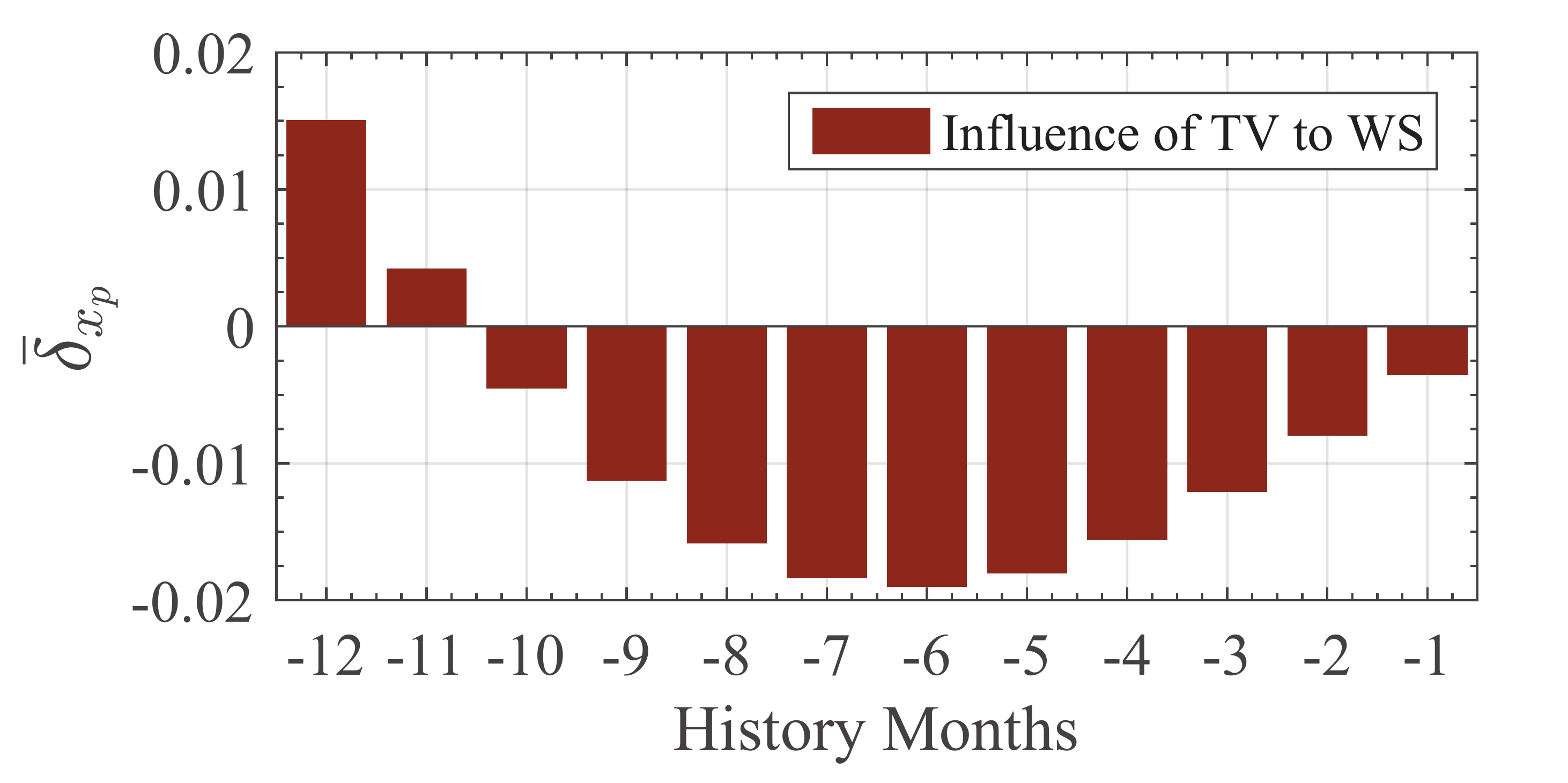}\label{fig:TV_influence}}
\subfigure[Fine-grained Volatility]{\includegraphics[width=0.5\columnwidth]{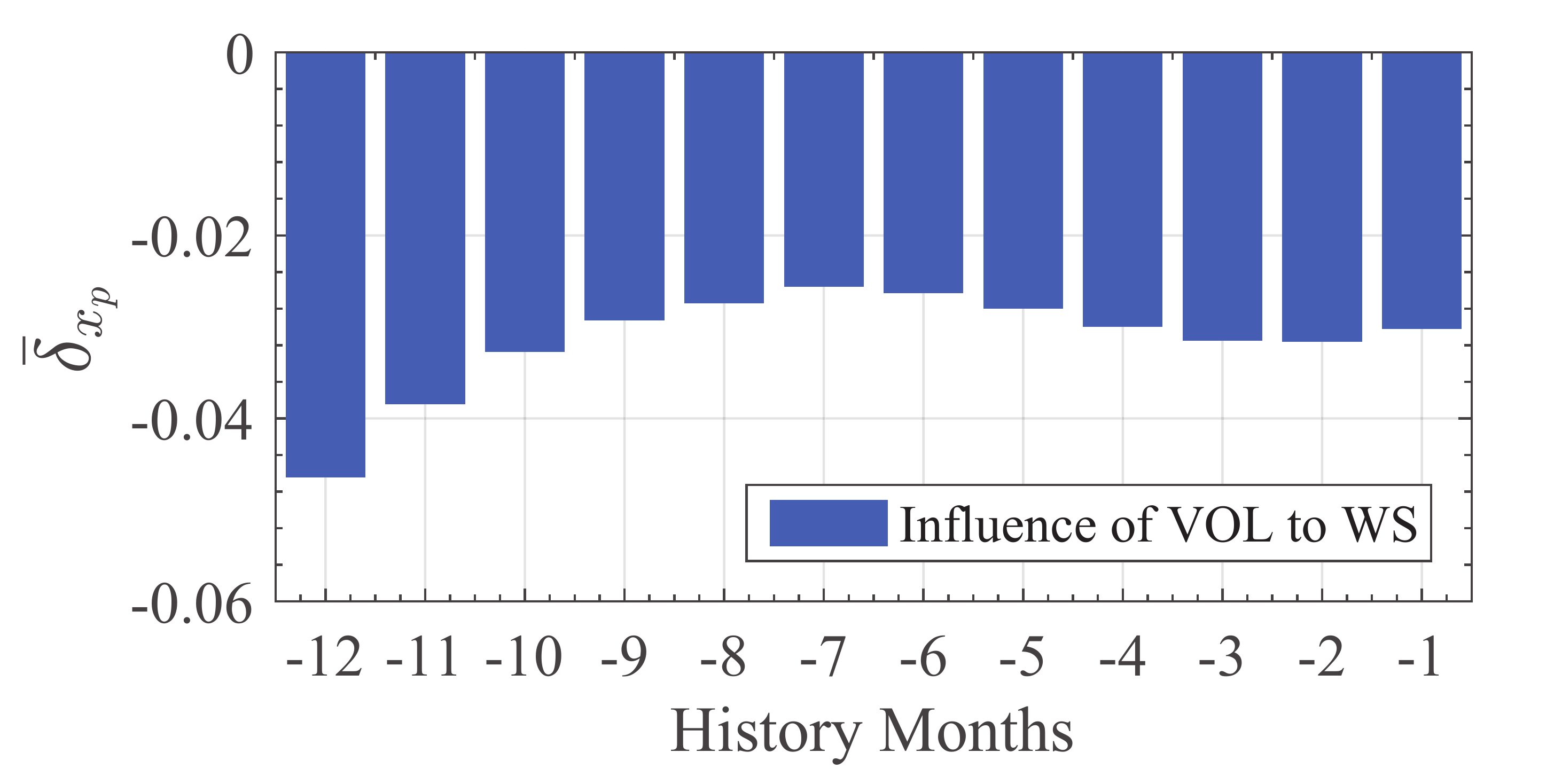}\label{fig:VOL_influence}}
\subfigure[Company Features]{\includegraphics[width=0.52\columnwidth]{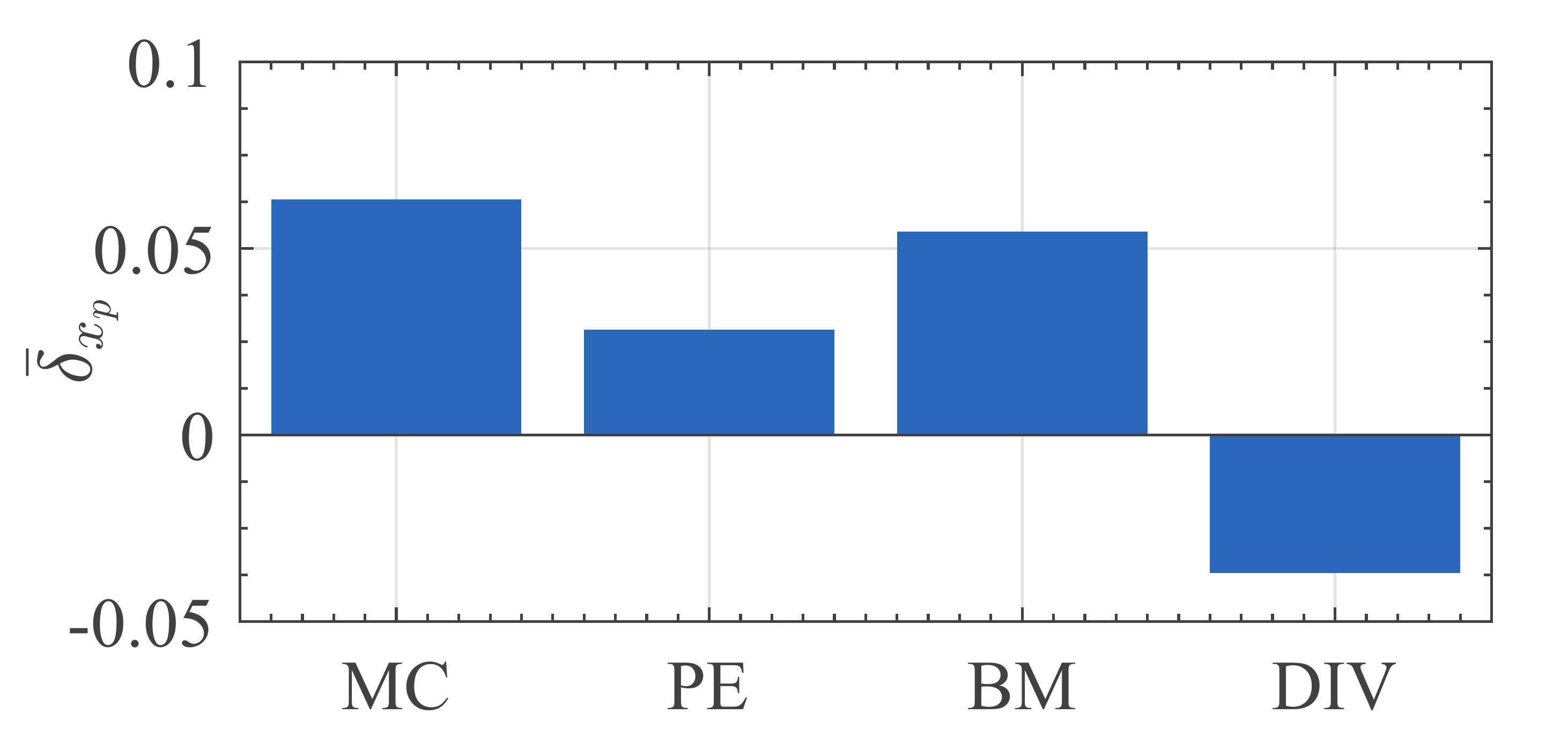}\label{fig:company_influence}}
\vspace{-0.3cm}
\caption{Influence of history trading features to winner scores.}\label{fig:market_influence}
\vspace{-0.1cm}
\end{figure*}

\subsection{Performance in Chinese Markets}
In order to further testify the robustness of our model, we run the back-test experiments of our model and baselines over the Chinese stock markets, which contains two exchanges: Shanghai Stock Exchange (SSE) and Shenzhen Stock Exchange (SZSE). The data are obtained from the WIND databese\footnote{\url{http://www.wind.com.cn/en/Default.html}}. The stocks are the RMB priced ordinary shares (A-share) and the total number of stocks used for experiment is 1,131. The time range of our data is from Jun. 2005 to Dec. 2018, with the period from Jun. 2005 -- Dec. 2011 used as the training/validation set and the rest as the test set. Since the Chinese markets cannot short sell, so we only use the $\bm{b}^+$ portfolio in the experiment.


The experimental results are given in Table~\ref{tab:metrics_CN}. From the table we can see that the performances of \name, \name-NP and \name-NC are better than that of other baselines again. This verifies the effectiveness of our model over the Chinese markets. By further comparing Table~\ref{tab:metrics_CN} with Table~\ref{tab:metrics}, it turns out that the risk of our model measured by AVOL and MDD in the Chinese markets is higher than that in the U.S. markets. This might be attributable to the market faultiness of emerging countries like China, with more speculative capital but less effective governance. The lack of short sell mechanism also contributes to the imbalance of market forces. The AVOL and MDD of the Market and other baselines in the Chinese markets are also higher than that in the U.S. markets. Compared with these baselines, the risk control ability of our model is still competitive. To sum up, the experimental results in Table~\ref{tab:metrics_CN} indicate the robustness of our model over emerging markets.

\begin{table}[t]\footnotesize
  \caption{Performance comparison on Chinese markets.}
  \label{tab:metrics_CN}\vspace{-0.3cm}
  \begin{tabular}{lcccccc}
    \toprule
   &{\bf APR}&{\bf \underline{AVOL}}&{\bf ASR}&{\bf \underline{MDD}}&{\bf CR}&{\bf DDR}\\    \midrule
    {\bf Market} & 0.037 & 0.260 & 0.141 & 0.595 & 0.062 & 0.135 \\
    {\bf TSM}    & 0.078 &0.420  & 0.186 & 0.533 & 0.147 & 0.225\\
    {\bf CSM}    & 0.023 & 0.392  & 0.058  & 0.633 & 0.036 & 0.064  \\
    {\bf RMR}    & 0.079 & 0.279 & 0.282  & 0.423  & 0.186  & 0.289 \\
    {\bf FDDR}   & 0.084 & 0.152  & 0.553 & 0.231 & 0.365 &  0.801\\
    \midrule
    {\bf AS-NC} & 0.104 & 0.113 & 0.916 &  { 0.163} &  0.648 & 1.103 \\
    {\bf AS-NP} & 0.122 & 0.105 & 1.163 &  { 0.136}  & { 0.895}  & 1.547 \\
    {\bf AS} & {\bf 0.125} & {\bf 0.103} & {\bf 1.220}  &  {\bf 0.135} &  {\bf 0.296} & {\bf 1.704}  \\
  \bottomrule
\end{tabular}\vspace{-0.2cm}
\end{table}

\subsection{Investment Strategies Interpretation}\label{sec:interpret_exp}


Here, we try to interpret the underlying investment strategies of \name, which is crucial for practitioners to better understanding this model. To this end, we use $\bar{\delta}_{x_p}$ in Eq.~\eqref{eq:delta_bar} to measure the influence of the stock features defined in Section~\ref{sec:features} to \name's winner selection. Figures~\ref{fig:PR_influence}-\ref{fig:TV_influence} plot the influences from the trading features. The vertical axis denotes the influence strengths indicated by $\bar{\delta}_{x_q}$, and the horizontal axis denotes how many months before the trading time. For example, the bar indexed by ``-12'' of the horizontal axis in Fig.~\ref{fig:PR_influence} denotes the influence of stock price rising rate (PR) at the time of twelve months ago.

As shown in Fig.~\ref{fig:PR_influence}, the influence of history price rising rate is heterogeneous along the time axis. The PR in long-term months, \ie 9 to 11 months ahead, has positive influence to winner scores, but for the short-term months, \ie 1 to 8 months ahead, the influence becomes negative. This result indicates that our model tends to buy the stocks with long-term rapid price increase (valid excellence) or with short-term rapid price retracement (over undervalued). This implies that \name behaviors like a long-term momentum but short-term reversion mixed strategy. Moreover, since price rising is usually accompanied by frequent stock trading, Fig.~\ref{fig:TV_influence} shows that the $\bar{\delta}_{x_p}$ of trading volumes (TV) has a similar tendency with the price rising rate (PR). Finally, as shown in Fig.~\ref{fig:VOL_influence}, the volatilities (VOL) have negative influence to winner scores for all history months. It means that our model trends to select low volatility stocks as winners, which indeed explains why \name can adapt to diverse market states.


Fig.~\ref{fig:company_influence} further exhibits the average influences of different company features to the winner score, \ie the $\bar{\delta}_{x_p}$ averaged on all history months. It turns out that Market Capitalization (MC), Price-earnings Ratio (PE), and  Book-to-market Ratio (BM) have positive influences. The three features are important company valuation factors for a listed company, which indicates that \name tends to select companies with sound fundamental values. In contrast, dividends mean a part of company values are returned to shareholders and could reduce the intrinsic value of a stock. That is why the influence of Dividends (DIV) is negative in our model.

To sum up, while \name is an AI-enabled investment strategy, the interpretation analysis proposed in Section~\ref{sec:interpret} can help to extract investment logics from \name. Specifically, \name suggests selecting the stocks as winners with {\em high long-term growth}, {\em low volatility}, {\em high intrinsic value}, and {\em being undervalued recently}.

\section{Related Works}
\label{sec:relate}

Our work is related to the following research directions.

{\bf Financial Investment Strategy:} Classic financial investment strategy includes Momentum, Mean Reversion, and Multi-factors. In the first work of BWSL~\cite{bwsl}, Jegadeesh and Titman found ``momentum'' could be used to select winners and losers. The momentum strategy buys assets that have had high returns over a past period as winners, and sells those that have had poor returns over the same period. Classic momentum strategies include the Cross Sectional Momentum (CSM)~\cite{jegadeesh2002cross} and the Time Series Momentum (TSM)~\cite{moskowitz2012time}. The mean reversion strategy~\cite{poterba1988mean} considers asset prices always return to their mean over a past period, so it buys assets with a price under their historical mean and sells above the historical mean. The multi-factor model~\cite{fama1996multifactor} uses factors to compute a valuation for each asset and buys/sells those assets with price under/above their valuations. Most of these financial investment strategies can only exploit a certain factor of financial markets and thus might fail in complex market environments.


{\bf Deep Learning in Finance:} In recent years, deep learning approaches begin to be applied in the financial areas. In the literature, L. Zhang \etal proposed to exploit frequency information to predict stock prices~\cite{hu2017state}. News and social media were used in price prediction in Refs.~\cite{news1,tweets}. 
Information about events and corporation relationships were used to predict stock prices in Ref.~\cite{event1,incorporating}. Most of these works focus on price prediction rather than end-to-end investment portfolio generation like us.


{\bf Reinforcement Learning in Finance:} The RL approaches used in investment strategies fall in two categories: the value-based and the policy-based~\cite{fischer2018reinforcement}. The value-based approaches learn a critic to describe the expected outcomes of markets to trading actions. Typical value-based approaches in investment strategies include Q-learning~\cite{Neuneier1995Optimal} and deep Q-learning~\cite{dql_stanford}. A defect of value-based approaches is the market environment is too complex to be approximated by a critic. Therefore, policy-based approaches are considered as more suitable to financial markets~\cite{fischer2018reinforcement}. The \name model also belongs to this category. A classic policy-based RL algorithm in investment strategy is the Recurrent Reinforcement Learning (RRL)~\cite{moody1998performance}. The FDDR~\cite{deng2017deep} model extends the RRL framework using deep neural networks. In the Investor-Imitator model~\cite{ding2018investor}, a policy-based deep RL framework was proposed to imitate the behaviors of different types of investors. Compared with RRL and its deep learning extensions, which focus on exploiting sequential dependence in financial signals, our \name model pays more attention to the interrelationships among assets. Moreover, deep RL approaches are often hard to deployed in real-life applications for unexplainable deep network structures. The interpretation tools offered by our model can solve this problem.

\section{Conclusions}
\label{sec:conclude}

In this paper, we proposed a RL-based deep attention network to design a BWSL strategy called \name. We also designed a sensitivity analysis method to interpret the investment logics of our model. Compared with existing RL-based investment strategies, \name fully exploits the interrelationship among stocks, and opens a door for solving the ``black box'' problem of using deep learning models in financial markets. The back-testing and simulation experiments over U.S. and Chinese stock markets showed that \name performed much better than other competing strategies. Interestingly, \name suggests buying stocks with high long-term growth, low volatility, high intrinsic value, and being undervalued recently. 

\section*{Acknowledgments}

J. Wang's work was partially supported by the National Natural Science Foundation of China (NSFC) (61572059, 61202426), the Science and Technology Project of Beijing (Z181100003518001), and the CETC Union Fund (6141B08080401). Y. Zhang's work was partially supported by the National Key Research and Development Program of China under Grant (2017YFC0820405) and the Fundamental Research Funds for the Central Universities. K. Tang's work was partially supported the National Social Sciences Foundation of China (No.14BJL028). J. Wu's work was partially supported by NSFC (71725002, 71531001, U1636210).

\bibliographystyle{ACM-Reference-Format}
\bibliography{BWSL}


\end{document}